\documentclass[pre,twocolumn,showpacs,superscriptaddress,preprintnumbers,floatfix]{revtex4}

\usepackage{dcolumn}
\usepackage{amsmath}
\usepackage{amssymb}
\usepackage{graphicx}
\usepackage{bm}   
\usepackage{bbm}   
\usepackage{verbatim}
\usepackage{stmaryrd}
\usepackage{amsthm}

\theoremstyle{break} 	\newtheorem{Cor}{Corollary}
\theoremstyle{plain} 	\newtheorem*{ProCor}{Proof}
\theoremstyle{break} 	\newtheorem{The}{Theorem}
\theoremstyle{plain} 	\newtheorem*{ProThe}{Proof}
\theoremstyle{break} 	\newtheorem{Prop}{Proposition}
\theoremstyle{plain} 	\newtheorem*{ProProp}{Proof}
\theoremstyle{plain}	
\theoremstyle{break}	\newtheorem*{Def}{Definition} 
\theoremstyle{plain}	\newtheorem*{Not}{Notation}
\theoremstyle{break}	\newtheorem*{Con}{Conjecture} 

\newcommand{\Prob}      {\mathrm{Pr}}
\newcommand{\Lang}      {\mathcal{L}}
\newcommand{\supp}      {\mathrm{supp}}
\newcommand{\subw}      {\mathrm{sub}}
\newcommand{\prcl}      {\mathcal{P}}

\newcommand{\mperiod} {\mathnormal{p}}
\newcommand{\Ldiv} {\mathcal{D}}
\begin{document}


\title{Language Diversity of\\
Measured Quantum Processes}

\author{Karoline Wiesner}
\email{wiesner@cse.ucdavis.edu}
\affiliation{Center for Computational Science \& Engineering and Physics Department,
University of California Davis, One Shields Avenue, Davis, CA 95616}
\affiliation{Santa Fe Institute, 1399 Hyde Park Road, Santa Fe, NM 87501}
\author{James P. Crutchfield}
\email{chaos@cse.ucdavis.edu}
\affiliation{Center for Computational Science \& Engineering and Physics Department,
University of California Davis, One Shields Avenue, Davis, CA 95616}
\affiliation{Santa Fe Institute, 1399 Hyde Park Road, Santa Fe, NM 87501}

\date{\today}

\bibliographystyle{unsrt}

\begin{abstract}
The behavior of a quantum system depends on how it is measured. How
much of what is observed comes from the structure of the quantum system
itself and how much from the observer's choice of measurement? We explore
these questions by analyzing the \emph{language diversity} of quantum
finite-state generators. One result is a new way to distinguish quantum
devices from their classical (stochastic) counterparts. While the diversity
of languages generated by these two computational classes is the same in
the case of periodic processes, quantum systems generally generate a 
wider range of languages than classical systems.
\end{abstract}

\pacs{
 03.67.Lx 
 03.67.-a 
 02.50.-r 
}
\preprint{Santa Fe Institute Working Paper 06-11-XXX}
\preprint{arxiv.org/quant-ph/0611XXX}

\maketitle




\section{Introduction}

Quantum computation has advanced dramatically from Feynman's initial
theoretical proposal \cite{feynman:82} to the experimental realizations
one finds today. The largest quantum device that has been implemented,
though, is a $7$ qubit register that can factor
a $3$ bit number \cite{knill:02} using Shor's algorithm \cite{shor:94}.
A review of this and other currently feasible quantum devices reveals
that, for now and the foreseeable future, they will remain small---in the
sense that a very limited number of qubits can be stored.
Far from implementing the theoretical ideal of a quantum Turing machine,
current experiments test quantum computation at the level of small
finite-state machines.

The diversity of quantum computing devices that lie between the extremes
of finite-state and (unbounded memory) Turing machines is substantially
less well understood
than, say, that for classical automata, as codified in the Chomsky hierarchy
\cite{hopcroft}. As an approach to filling in a quantum hierarchy, comparisons
between classical and quantum automata can be quite instructive. 

Such results are found for automata at the level of finite-state machines
\cite{kondacs:97,moore:00,wiesner:06}. For example, the regular languages
are recognized by finite-state machines (by definition), but quantum
finite-state machines, as defined in Ref. \cite{moore:00}, cannot recognize
all regular languages. This does not mean, however, that quantum automata
are strictly less powerful than their classical counterparts. There are
nonregular languages that are recognized by quantum finite-state machines
\cite{bertoni:01}. These first results serve to illustrate the need for
more work, if we are to fully appreciate the properties of quantum devices
even at the lowest level of some presumed future quantum computational
hierarchy.

The comparison of quantum and classical automata has recently been extended
to the probabilistic languages recognized by stochastic and quantum finite-state
machines \cite{wiesner:06}. There, quantum finite-state generators
were introduced as models of the behaviors produced by quantum systems and as
tools with which to quantify their information storage and processing
capacities.

Here we continue the effort to quantify information processing in simple
quantum automata. We will show how a quantum system's possible behaviors
can be characterized by the diversity of languages it generates under different
measurement protocols. We also show how this can be adapted to measurements,
suitably defined, for classical automata. It turns out that the diversity of
languages, under varying measurement protocols, provides a useful way to
explore how classical and quantum devices differ. A measured quantum system
and its associated measured classical system can generate rather different
sets of stochastic languages. For periodic processes, the language diversities
are the same between the quantum and counterpart classical systems. However,
for aperiodic processes quantum systems are more diverse, in this
sense, and potentially more capable.

In the following, we first review formal language and automata theory,
including stochastic languages, stochastic and quantum finite-state
generators, and the connection between languages and behavior. We then
introduce the \emph{language diversity} of a finite-state automaton
and analyze a number of example processes, comparing quantum and
classical models. We conclude with a few summary remarks and contrast
the language diversity with \emph{transient information}, which measures
the amount of information an observer needs to extract in order to predict
which internal state a process is in \cite{crutchfield:03}.


\section{Formal Languages and Behavior}
\label{sec:language}

Our use of formal language theory differs from most in how it analyzes
the connection between a language and the systems that can generate
it. In brief, we observe a system through a finite-resolution measuring
instrument, representing each measurement with a \emph{symbol} $\sigma$ from
discrete \emph{alphabet} $\Sigma$. The temporal behavior of a system, then, is
a string or a \emph{word} consisting of a succession of measurement symbols.
The collection of all (and only) those words is the \emph{language} that
captures the possible, temporal behaviors of the system.

\begin{Def}
A \emph{formal language} $\mathcal{L}$ is a set of \emph{words}
$w = \sigma_0\sigma_1 \sigma_2  \ldots$ each of which consists of a series of symbols
$\sigma_i \in \Sigma$ from a discrete alphabet $\Sigma$. 
\end{Def}

$\Sigma^*$ denotes the set of all
possible words of any length formed using symbols in $\Sigma$. We denote a
word of length $L$ by $\sigma^L = \sigma_0 \sigma_1 \ldots \sigma_{L-1}$, with
$\sigma_i \in \Sigma$. The set of all words of length $L$ is $\Sigma^L$.

Since a formal language, as we use the term, is a set of observed words
generated by a process, then each \emph{subword}
$\sigma_i \sigma_{i+1} \ldots \sigma_{j-1} \sigma_j, i \leq j,~ i,j = 0,1,
\ldots, L-1$ of a word $\sigma^L$ has also been observed and is
considered part of the language. This leads to the following definition.

\begin{Def}
A language $\mathcal{L}$ is \emph{subword closed} if, for each
$w \in \Lang$, all of $w$'s subwords $\mathrm{sub}(w)$ are also
members of $\Lang$: $\subw(w) \subseteq \mathcal{L}$.
\end{Def}

Beyond a formal language listing which words (or behaviors) occur and
which do not, we are also interested in the probability of their
occurrence. Let $\Prob(w)$ denote the probability of word $w$, then
we have the following definition.

\begin{Def}
A \emph{stochastic language} $\Lang$ is a formal language with a
\emph{word distribution} $\Prob(w)$ that is normalized at each length $L$:
\begin{equation}
\sum_{\{\sigma^L \in \mathcal{L}\}} \Prob(\sigma^L) = 1 ~,
\end{equation}
with $0 \leq \Prob(\sigma^L) \leq 1$.
\end{Def}

\begin{Def}
Two stochastic languages $\Lang_1$ and $\Lang_2$ are said to be
$\delta$-similar if 
$\forall \sigma^L \in \Lang_1$ and $\sigma^{\prime L} \in \Lang_2$ : $\vert
\Prob(\sigma^L)-\Prob(\sigma^{\prime L}) \vert \leq \delta$, for all $L$ and a
specified $0 \leq \delta \leq 1$. If this is true for $\delta = 0$, then the
languages are equivalent.
\end{Def}

For purposes of comparison between various computational models, it is
helpful to refer directly to the set of words in a stochastic language
$\Lang$. This is the \emph{support} of a stochastic language:
\begin{equation}
\supp(\Lang) = \{w \in \Lang: ~\Prob(w) > 0 \} ~.
\end{equation}

The support itself is a formal language. Whenever we compare formal and
stochastic languages we add the respective subscripts and write
$\Lang_{formal}$ and $\Lang_{stoch}$.

\section{Stochastic Finite-State Generators}
\label{sec:sfm}

Automata with finite memory---\emph{finite-state machines}---consist of a
finite set of states and transitions between them \cite{hopcroft}.
Typically, they are used as \emph{recognition} devices, whereas we are
interested in the generation of words in a stochastic language. So here
we will review models for classical and quantum generation, referring the
reader to Ref. \cite{paz} for details on recognizers and automata
in general.

\begin{Def}
\cite{wiesner:06}
\label{def:mea}
A \emph{stochastic generator} $G$ is a tuple
$\{ S,Y, \{ T(y) \} \}$ where
\begin{enumerate}
\item $S$ is a finite set of states, with $|S|$ denoting its cardinality.
\item $Y$ is a finite alphabet for output symbols.
\item $\{T(y), y \in Y \}$ is a set of $|Y|$ square stochastic matrices of
	order $|S|$. $|Y|$ is the cardinality of $Y$, the components $T_{ij}(y)$
	give the probability of moving to state $s_j$ and emitting $y$ when in state
	$s_i$.
\item At each step a symbol 
	$y \in Y$ is emitted and the machine updates its state. Thus,
	$\sum_{y\in Y}\sum_{j} T_{ij}(y) = 1$.
\end{enumerate}
\end{Def}

\begin{Def}
\label{def:dg}
A \emph{deterministic generator} ($DG$) is a $G$ in which each matrix
$T(y)$ has at most one nonzero entry per row.
\end{Def}

\subsection{Process languages}

\begin{Def}
A \emph{process language} ${\prcl}$ is a stochastic language that is
subword closed.
\end{Def}

The output of a stochastic generator (as well as the quantum generator
introduced below) is a process language; for the proof see Ref.
\cite{wiesner:06}. Thus, all stochastic languages discussed in the following
are process languages.

\begin{Def}
A \emph{periodic process language} with period $N$ is a process language such
that $\forall w =\sigma_0\sigma_1\dots \sigma_n \in \prcl$ with $n\geq N$:
$\sigma_{i} = \sigma_{i+N}$.
\end{Def}

Before discussing the languages associated with a $G$, we must introduce some
helpful notation.

\begin{Not}
Let $| \eta \rangle = (1 1 \ldots 1 1)^T$ denote a column vector with
$|S|$ components that are all $1$s.
\end{Not}

\begin{Not}
The \emph{state vector} $\langle\pi | = (\pi_0, \pi_1, \ldots, \pi_{|S|-1})$
is a row vector whose components, $0 \leq \pi_i \leq 1$, give the
probability of being in state $s_i$. The state vector is normalized in
probability: $\sum_{i=0}^{|S|-1} \pi_i = 1$. The
\emph{initial state distribution} is denoted $\langle\pi^0\vert$.
\end{Not}

The state-to-state transition probabilities of a $G$, independent of 
outputs, are given by the \emph{state-to-state transition matrix}:
\begin{equation}
\label{eqn:t}
T = \sum_{y\in Y} T(y)~, 
\end{equation}
which is a stochastic matrix: i.e.,
$0\leq T_{ij} \leq 1$ and $\sum_j T_{ij} = 1$.

The generator updates its state distribution after each time step as follows:
\begin{equation}
\langle \pi^{t+1} | = \langle \pi^t | T(y) ~, 
\end{equation}
where (re)normalization of the state vector is assumed.

If a $G$ starts in state distribution $\langle\pi^0|$, the probability of
generating $y^L$ is given by the state vector without renormalization
\begin{equation}
\label{eqn:cpryL}
\Prob(y^L) = \langle \pi^0 |  T(y^L) | \eta \rangle~,
\end{equation}
where $T(y^L) = \prod_{i=0}^{L-1} T(y_i)$ represents the
assumption in our model that all states are accepting. This, in turn, is a
consequence of our focusing on process languages, which are subword closed.


\section{Quantum Generators}

Quantum generators are a subset of \emph{quantum machines} (or
\emph{transducers}), as defined
in Ref. \cite{wiesner:06}. Their architecture consists of a set of internal
states and transitions and an output alphabet that labels transitions. For
simplicity here we focus on the definition of generators, without repeating
the general definition of quantum transducers. Our basic
\emph{quantum generator} ($QG$) is defined as follows.

\begin{Def}
\cite{wiesner:06}
\label{def:qa}
A  $QG$ is a tuple
$\{Q,\mathcal{H},Y,\mathbf{T}(Y)\} \}$ where
\begin{enumerate}
\item $Q = \{q_i: i = 0, \ldots, n-1 \}$ is a set of $n = |Q|$
	\emph{internal states}.
\item The \emph{state space} ${\mathcal H}$ is an $n$-dimensional Hilbert space.
\item The \emph{state vector} is $\langle\psi\vert \in \mathcal{H}$.
\item $Y$ is a finite alphabet for output symbols. $\lambda \notin Y$ denotes the
	null symbol.
\item $\mathbf{T}(Y)$ is a set of $n$-dimensional \emph{transition matrices}
	$\{T(y) = P(y)\cdot U, y \in Y \}$ that are products of a
	unitary matrix $U$ and a projection operator $P(y)$ where
\begin{enumerate}
\item $U$ is an  $n$-dimensional unitary 
	\emph{evolution operator} that governs the evolution of the state vector. 
\item $\mathbf{P}(Y)$ is a set of $n$-dimensional
	\emph{projection operators}---$\mathbf{P} = \{ P(y) : y \in Y \cup \{\lambda\} \}$---that
	determines how a state vector is measured.The $P(y)$ are Hermitian matrices.
\end{enumerate}
\end{enumerate}
At each time step a $QG$ outputs a symbol $y \in Y$ or the null symbol $\lambda$
and updates its state vector.
\end{Def}

The output symbol $y$ is identified with the measurement outcome. The symbol
$\lambda$ represents the event of no measurement.
In the following we will concentrate on deterministic quantum generators.
They are more transparent than general (nondeterministic) QGs, but still
serve to illustrate the relative power of quantum and classical generators.

\begin{Def}
\label{def:qdg}
A \emph{quantum deterministic generator} ($QDG$) is a $QG$ in which each matrix
$T(y)$ has at most one nonzero entry per row.
\end{Def}

\subsection{Observation and Operation}
\label{sec:qfm_observe}

The projection operators determine how output symbols are
generated from the internal, hidden dynamics. In fact, the only
way to observe a quantum process is to apply a projection operator to the
current state. In contrast with classical processes, the measurement event
disturbs the internal dynamics. The projection operators are familiar from quantum
mechanics and can be defined in terms of the internal states as follows.

\begin{Def}
A \emph{projection operator} $P(y)$ is the linear operator
\begin{equation}
P(y) = \sum_{\kappa \in {\mathcal H_{y}}} |\phi_{\kappa}\rangle
\langle\phi_{\kappa}|~,
\end{equation}
where $\kappa$ runs over the indices of a one- or higher-dimensional
subspace ${\mathcal H_{y}}$ of the Hilbert space and the
$\phi_{\kappa}$ span these subspaces.
\end{Def}

We can now describe a $QG$'s operation. $U_{ij}$ is the transition amplitude
from state $q_i$ to state $q_j$. Starting in state $\langle\psi_0\vert$ the
generator updates its state by applying the unitary matrix $U$. Then the
state vector is projected using $P(y)$ and renormalized. Finally, symbol
$y \in Y$ is emitted. In other words, a single time-step of a $QG$ is given by:
\begin{equation}
\langle \psi(y) \vert = \langle\psi^0\vert U  P(y)
  ~,
\label{eqn:psiy}
\end{equation}
where (re)normalization of the state vector is assumed. The state vector after
$L$ time steps when emitting string $y^L$ is

\begin{equation}
\langle \psi(y^L) \vert   
  = \langle\psi^0\vert \prod_{i=0}^{L-1} \left(U P(y_i)\right) ~.
\label{eqn:psiyL}
\end{equation}

We can now calculate symbol and word probabilities of the process language
generated by a $QG$. Starting the $QG$ in $\langle\psi^0\vert$ the
probability of output symbol $y$ is given by the state vector without
renormalization:
\begin{equation}
\label{eqn:qpry}
\Prob(y) =  \left\Vert \psi(y) \right\Vert^2 ~.
\end{equation}
By extension, the probability of output string $y^L$  is
\begin{equation}
\label{eqn:qpryL}
\Prob(y^L) =  \left\Vert \psi(y^L) \right\Vert^2 ~.
\end{equation}

\subsection{Properties}
\label{sec:qfm_prop}

In Ref. \cite{wiesner:06} we established a number of properties of $QG$s:
their consistency with quantum mechanics, that they generate process
languages, and their relation to stochastic generators and to quantum and
stochastic recognizers. Here we avail ourselves of one property in particular
of $QDG$s---for a given $QDG$ there is always an equivalent (classical)
deterministic generator. The latter is obtained by squaring the matrix
elements of the $QDG$'s unitary matrix and using the same projection
operators. The resulting state-to-state transition matrix is doubly
stochastic; i.e., $0\leq T_{ij} \leq 1$ and $\sum_i
T_{ij} = \sum_j T_{ij}= 1$. 

\begin{The}
\label{the:DG}
Every process language generated by a $QDG$ is generated by some $DG$.
\end{The}

\begin{ProThe}
See Ref. \cite{wiesner:06}.
\end{ProThe}

This suggests that the process languages generated by $QDG$s are a subset
of those generated by $DG$s. In the following, we will take a slightly
different perspective and ask what set of languages a given $QDG$
can generate as one varies the \emph{measurement protocol}---that is, the
choice of measurements.

\section{Language diversity}

The notion of a measurement protocol is familiar from quantum mechanics:
We define the \emph{measurement period} as the number of applications of
a projection operator relative to the unitary evolution time step. For a
classical system this is less familiar, but it will be used in the same way.
The measurement period here is the period of observing an output symbol
relative to the internal state transitions. The internal dynamics remain
unaltered in the classical case, whether the system is measured or not. In the
quantum case, as is well known, the situation is quite different. Applying a
projection operator disturbs the internal dynamics.

\begin{Def}
A process observed with \emph{measurement period} $\mperiod$ is measured every
$\mperiod$ time steps.
\end{Def}

Note that this model of a measurement protocol, by which we subsample the
output time series, is related to von Mises version of probability theory
based on ``collectives'' \cite{howson:95}.

The resulting observed behavior can be described in terms of the state-to-state
transition matrix and the projection operators. For a classical finite-state
machine this is:
\begin{equation}
\label{eqn:cfreq}
\langle \pi(y)^{t+\mperiod} \vert = \langle \pi^t \vert T^{\mperiod-1} T(y)~,
\end{equation}
where $\langle\pi(y)^{t+\mperiod}|$ is the state distribution vector after
$\mperiod$ time steps and after observing symbol $y$. Note that
$T(y) = T P(y)$.

For a quantum finite-state machine we have, instead:
\begin{equation}
\label{eqn:qfreq}
\langle\psi(y)^{t+\mperiod}\vert = \langle\psi^t\vert U^{\mperiod} P(y)~.
\end{equation}
In both cases we dropped the renormalization factor.

The stochastic language generated by a particular quantum finite-state
generator $G$ for a particular measurement period $\mperiod$ is labeled
$\Lang^{\mperiod}(G)$. Consider now the set of languages generated by $G$ for
varying measurement period $\{\Lang^{\mperiod}(G)\}$.

\begin{Def}
The \emph{language diversity} of a  (quantum or classical) finite-state
machine $G$ is the logarithm of the total number
$\vert \{\Lang^\mperiod(G)\} \vert$ of
stochastic languages that $G$ generates as a function of measurement
period $\mperiod$:
\begin{equation}
\Ldiv (G) = \log_2 \vert \{ \Lang^\mperiod(G) \} \vert ~.
\end{equation}
\end{Def}

Whenever we are interested in comparing the diversity in terms of formal
and stochastic languages we add the respective subscript and write
$\Ldiv_{formal}(G)$ and $\Ldiv_{stoch}(G)$, respectively. Here, 
$\Ldiv_{formal} = log_2 \vert \Lang^{\mperiod}_{formal}\vert $.
In general, $\Ldiv_{stoch}(G) > \Ldiv_{formal}(G)$ for any particular $G$.

In the following we will demonstrate several properties related to the
language diversity of classical and quantum finite-state machines. 

Since every $\Lang(QDG)$ is generated by some $DG$, at first blush
one might conclude that $DG$s are at least as powerful as $QDG$s. However,
as pointed out in Ref. \cite{wiesner:06}, this is true only for one
particular measurement period. In the following examples we will study
the dependence of the generated languages on the measurement period.
It will become clear that Theorem~\ref{the:DG} does not capture all of
the properties of a $QDG$ and its classical analog $DG$. For all but the
periodic processes of the following examples the \emph{language diversity} is
larger for the $QDG$ than its $DG$ analog, even though the projection
operators are identical.

These observations suggest the following.
\begin{Con}
$\Ldiv(QDG) \geq \Ldiv(DG)$.
\end{Con}

The inequality becomes an equality in one case.
\begin{Prop}
\label{the:perstochlang}
For a $QDG$ $G$  generating a periodic stochastic language $\Lang$ and its
analog $DG$ $G^{\prime}$ 
\begin{equation}
\Ldiv(G) = \Ldiv(G^{\prime})~.
\end{equation}
\end{Prop}

\begin{ProProp}
For any measurement period $\mperiod$ and word length $L$ words
$y^L \in \Lang(G)$ and
$y^{\prime L} \in \Lang(G^{\prime})$ with $y^L = y^{\prime L}$
have the same probability:
$\Prob(y^L) = \Prob(y^{\prime L})$. That is,
\begin{equation*}
\Prob(y^L) = \Vert \psi^0 U^{\mperiod}P{(y_0)}U^{\mperiod}P{(y_1)}\dots
  U^{\mperiod}P{(y_{L-1})} \Vert^2
\end{equation*}
and
\begin{equation*}
\Prob(y^{\prime L}) = \langle \pi^0 |
  T^{\mperiod}P{(y_0)}T^{\mperiod}P{(y_1)}\dots
  T^{\mperiod}P{(y_{L-1})} | \eta \rangle ~.
\end{equation*}
Due to determinism and periodicity $\Prob(y^L) = 0$ or $1$, and also
$\Prob(y^{\prime L}) = 0$ or $1$ for all possible $\psi^0$ and $\pi^0$,
respectively.  Since $U=T$, the probabilities are equal. $_{\square}$
\end{ProProp}

We can give an upper bound for $\Ldiv$ in this case.

\begin{Prop}
\label{the:upperboundPQG}
For a $QG$ $G$ generating a periodic process language $\Lang$ with period
$N$:
\begin{equation}
\Ldiv(G) \leq log_2(|Y|+N(N-1)) ~.
\end{equation}
\end{Prop}

\begin{ProProp}
Since $\Lang(G)$ is periodic, $\Lang^{\mperiod}(G) = \Lang^{\mperiod + N}(G)$.
For $\mperiod = N, 2N, \dots$: $\Lang^p(G) = \{y^*\}$, $y\in Y$.  For
$\mperiod = N+i, 2N+i, \dots,\, 0<i<N$: $\Lang^p(G) =
\subw((\sigma_0\sigma_1\dots\sigma_{N-1})^*)$ and all its cyclic permutations
are generated, in total $N$ for each $p$. This establishes an upper bound of
$|Y|+N(N-1)$.
\end{ProProp}

For general quantum processes there exists an upper bound for the
language diversity.
\begin{Prop}
\label{the:upperboundQG}
For a $QGD$ $G$
\begin{equation}
\Ldiv(G) \leq log_2(|Y|+k(k-1)) ~,
\end{equation}
where $k$ is the integer giving
\begin{equation}
U^{k} = I+\iota J~,
\end{equation}
$I$ is the identity matrix, $\iota \ll 1$, and $J$ is a diagonal matrix
$\sum_i\vert J_{ii}\vert^2 \leq 1$.
\end{Prop}

\begin{ProProp}
It was shown in Ref. \cite{moore:00} (Thms. 6 and 7), that any $n\times n$
unitary  $U$ can be considered as rotating an $n-$dimensional torus. Then for
some $k$ $U^k$ is within a small distance of the identity matrix. Thus, $k$
can be considered the \emph{pseudo-period} of the process, compared to a
strictly periodic process with \emph{period} $N$ and $U^N = I$.

Thus, $\Lang^{\mperiod}(G)$ and $\Lang^{\mperiod + k}(G)$ are $\delta$-similar
with $\delta \ll 1$. For
$\mperiod = k:$ $U^{\mperiod} = I +\iota J$, generating $\Lang = \{y^*\}$. Using the
same argument as in the proof of Prop.~\ref{the:upperboundPQG} to lower the
bound by $k$ this establishes the upper bound for $\Ldiv(G)$.$_{\square}$
\end{ProProp}

It should be noted that the upper bound on $\Ldiv$  depends on the parameter
$\delta$ defining the similarity of languages $\Lang^{\mperiod}(G)$ and
$\Lang^{\mperiod + k}(G)$. In general, the smaller $\delta$ is, the larger is $k$.

\begin{Prop}
\label{pro:formlang}
For a $QDG$ $G$ generating a periodic process language the number of formal languages
$|\Lang_{formal}(G)|$ equals the number of stochastic languages
$|\Lang_{stoch}(G)|$
\begin{equation}
\Ldiv_{formal}(G) = \Ldiv_{stoch}(G).
\end{equation}
\end{Prop}

\begin{ProProp}
It is easily seen that any $QG$ generating a periodic process is
deterministic: its unitary matrix has only $0$ and $1$ entries. It
follows that word probabilities are either $0$ or $1$ and so there
is a one-to-one mapping between the stochastic language generated
and the corresponding formal language.$_{\square}$
\end{ProProp}

\begin{Cor}
For a $QDG$ $G$ generating a  periodic process and its analog $DG$
$G^{\prime}$:
\begin{equation}
\Ldiv_{formal}(G) = \Ldiv_{formal}(G^{\prime}) =
\Ldiv_{stoch}(G) = \Ldiv_{stoch}(G^{\prime}) ~.
\end{equation}
\end{Cor}

\begin{ProCor}
The Corollary follows from Prop.~\ref{the:perstochlang} and
a straightforward extension of Proposition~\ref{pro:formlang}
to classical periodic processes.$_{\square}$
\end{ProCor}

\section{Examples}

The first two examples, the iterated beam splitter and the quantum kicked top,
are quantum dynamical systems that are observed using complete measurements.
In quantum mechanics, a \emph{complete measurement} is defined as a
nondegenerate measurement operator, i.e., one with nondegenerate eigenvalues.
The third example, the distinct period-5 processes, illustrates processes
observed via incomplete measurements. Deterministic quantum and stochastic
finite-state generators are constructed and compared for each example.

\subsection{Iterated beam splitter}

The iterated beam splitter is a simple quantum process, consisting
of a photon that repeatedly passes through a loop of beam splitters and
detectors, with one detector between each pair of beam splitters
\cite{wiesner:06}. Thus, as the photon traverses between one beam splitter
and the next, its location in the upper or lower path between them
is measured nondestructively by the detectors. The resulting output sequence
consists of symbols $0$ (upper path) and $1$ (lower path).

The operators have the following matrix representation in the
experiment's eigenbasis:
\begin{eqnarray}
\label{eqn:had}
U = \frac{1}{\sqrt{2}}
	\left(\begin{array}{cc} 1 & 1 \\ 1 & -1 \end{array}\right) ~,\\
P(0) = 	\left(\begin{array}{cc} 1 & 0 \\ 0 & 0 \end{array}\right) ~,\\
P(1) = 	\left(\begin{array}{cc} 0 & 0 \\ 0 & 1 \end{array}\right) ~.
\end{eqnarray}

Observing with different measurement periods, the generated language varies
substantially. As can be easily seen with Eqs.~(\ref{eqn:qpryL}) and
(\ref{eqn:qfreq}), three (and only three) languages are generated as one
varies $\mperiod$. They are summarized in Table~\ref{tab:ibs} for all
$y^L\in\Lang$ and for $n=0,1,2\dots$, which is used to parametrize the
measurement period. The language diversity of the $QDG$ is then $\Ldiv =
\log_2(3)$. We can compare this to the upper bound given in
Prop.~\ref{the:upperboundQG}. In the case of the unitary matrix $U$
given above $k=2$, since $UU=I$. $U$ is also known as the \emph{Hadamard}
matrix. Thus, the upper bound for the language diversity in this
case is $\Ldiv \leq log_2(4)$.

\begin{table}[tbp]
\begin{tabular}{|l|c|c|c|c|}
\hline
\multicolumn{5}{|c|}{Iterated Beam Splitter Language Diversity} \\
\hline
Machine & $\mperiod$ & $\supp(\Lang)$ & $\Lang$ & $\Ldiv$\\
Type & & & & \\
\hline
$QDG$  & $2n$   & $(0+1)^*$  & $\Prob(y^L) = 2^{-L}$ & \\
       & $2n+1$ & $0^*$      & $\Prob(y^L) = 1$ & \\
       & $2n+1$ & $1^*$      & $\Prob(y^L) = 1$ & $1.58$\\
\hline
$DG$   & $n$   & $(0+1)^*$  & $\Prob(y^L) = 2^{-L}$ & $0$ \\
\hline
\end{tabular}
\caption{Process languages generated by the $QDG$ for the iterated
  beam splitter and by the classical $DG$.
  The measurement period takes a parameter $n = 0, 1, 2 \dots$.
  The word probability is given for all $y^L\in\Lang$.
  }
\label{tab:ibs}
\end{table}

The classical equivalent $DG$ for the iterated beam splitter,
constructed as described in Ref. \cite{wiesner:06}, is given by the
following state-to-state transition matrix: 
\begin{eqnarray*}
T =
	\left(\begin{array}{cc}  \frac{1}{2}& \frac{1}{2} \\
 \frac{1}{2}& \frac{1}{2} \end{array}\right) ~.
\end{eqnarray*}
Using Eqs.~(\ref{eqn:cpryL}) and (\ref{eqn:cfreq}), we see that only
one language is generated for all $\mperiod$. This is the language of the
\emph{fair coin} process, a random sequence of $0$s and $1$s, see
Table~\ref{tab:ibs}. Thus, $\Ldiv (DG) = 0$.

\subsection{Quantum kicked top}

The periodically kicked top is a familiar example of a finite-dimensional
quantum system whose classical limit exhibits various degrees of chaotic
behavior as a function of its control parameters \cite{haake:87}. For a
spin-$1/2$ system the unitary matrix is:
\begin{eqnarray*}
U =
	\left(\begin{array}{cc} \frac{1}{\sqrt{2}} &-\frac{1}{\sqrt{2}}  \\
\frac{1}{\sqrt{2}} & \frac{1}{\sqrt{2}} \end{array}\right) 
\cdot 	\left(\begin{array}{cc} e^{-ik} & 0 \\ 0 & e^{-ik} \end{array}\right)
\end{eqnarray*}
and the projection operators are:
\begin{align*}
P(0) & = \left( \begin{array}{cc} 1 & 0 \\ 0 & 0 \end{array}\right) ~, \\
P(1) & = \left( \begin{array}{cc} 0 & 0 \\ 0 & 1 \end{array}\right)~.
\end{align*}

Since this $QDG$ $G$ is deterministic, its classical $DG$ $G^{\prime}$ exists and is given by:
\begin{eqnarray*}
T =
	\left(\begin{array}{cc}  \frac{1}{2}& \frac{1}{2} \\
 \frac{1}{2}& \frac{1}{2} \end{array}\right) ~.
\end{eqnarray*}
The process languages generated by this $QDG$ and its analog $DG$ are given in
Table~\ref{tab:qkt}. 
The language diversity is $\Ldiv (G) = \log_2 (5)$.
Whereas the language diversity of classical counterpart $DG$ is
$\Ldiv (G^{\prime}) = 0$, since it generates only the language of the fair
coin process.

\begin{table}[tbp]

\begin{tabular}{|l|c|c|c|c|}
\hline
\multicolumn{5}{|c|}{Spin-$1/2$ Quantum Kicked Top Language Diversity} \\
\hline
Machine & $\mperiod$ & $\supp(\Lang)$ & $\Lang$ & $\Ldiv$\\
Type & & & & \\
\hline
$QDG$ & $4n+1, 4n+3$ & $(0+1)^*$        & $\Prob(y^L) = 2^{-L}$ & \\
      & $4n+2$       & $\subw((01)^*)$  & $\Prob(((01)^*)^L) = 1/2$ & \\
      &              &                  & $\Prob(((10)^*)^L) = 1/2$ & \\
      & $4n+2$       & $\subw((10)^*)$  & $\Prob(((10)^*)^L) = 1/2$ & \\
      &              &                  & $\Prob(((01)^*)^L) = 1/2$ & \\
      & $4n$         & $0^*$            & $\Prob(y^L) = 1$ & \\
      & $4n$         & $1^*$            & $\Prob(y^L) = 1$ & $2.32$\\
\hline
$DG$  & $n$          & $(0+1)^*$        & $\Prob(y^L) = 2^{-L}$ & $0$\\
\hline
\end{tabular}
\caption{\label{tab:qkt}Process languages generated by the $QDG$ for the spin-$1/2$
  quantum kicked top and its corresponding classical $DG$.
  The measurement period, again, is parametrized by $n=0,1,2\dots$.
  The word probability is given for all $y^L\in\Lang$.
  }
\end{table}

\subsection{Period-5 process}

As examples of periodic behavior and, in particular, of incomplete measurements,
consider the binary period-$5$
processes distinct up to permutations and ($0\leftrightarrow 1$)
exchange. There are only three such processes: $(11000)^*$, $(10101)^*$, and 
$(10000)^*$ \cite{Feld02a}. They all have the same state-to-state transition
matrix---a period-5 permutation. This irreducible, doubly stochastic
matrix is responsible for the fact that the $QDG$ of a periodic process and
its classical $DG$ have the same properties. Their state-to-state unitary
transition matrix is given by
\begin{equation}
T = U = \left(
\begin{array}{ccccc}
0 & 0 & 1 & 0 & 0 \\
0 & 0 & 0 & 1 & 0 \\
0 & 1 & 0 & 0 & 0 \\
0 & 0 & 0 & 0 & 1 \\
1 & 0 & 0 & 0 & 0 \\
\end{array}\right)~.
\end{equation}

The projection operators  differ between the processes with different template
words, of course. For template word $10000$, they are:
\begin{align}
P(0) & = \left(
\begin{array}{ccccc}
1 & 0 & 0 & 0 & 0 \\
0 & 1 & 0 & 0 & 0 \\
0 & 0 & 0 & 0 & 0 \\
0 & 0 & 0 & 1 & 0 \\
0 & 0 & 0 & 0 & 1 \\
\end{array}\right)~, \\
P(1) & = \left(
\begin{array}{ccccc}
0 & 0 & 0 & 0 & 0 \\
0 & 0 & 0 & 0 & 0 \\
0 & 0 & 1 & 0 & 0 \\
0 & 0 & 0 & 0 & 0 \\
0 & 0 & 0 & 0 & 0 \\
\end{array}\right)~.
\end{align}
For $11000$, they are:
\begin{align}
P(0) & = \left(
\begin{array}{ccccc}
1 & 0 & 0 & 0 & 0 \\
0 & 0 & 0 & 0 & 0 \\
0 & 0 & 0 & 0 & 0 \\
0 & 0 & 0 & 1 & 0 \\
0 & 0 & 0 & 0 & 1 \\
\end{array}\right)~, \\
P(1) & = \left(
\begin{array}{ccccc}
0 & 0 & 0 & 0 & 0 \\
0 & 1 & 0 & 0 & 0 \\
0 & 0 & 1 & 0 & 0 \\
0 & 0 & 0 & 0 & 0 \\
0 & 0 & 0 & 0 & 0 \\
\end{array}\right)~.
\end{align}
And for word $10101$, they are:
\begin{align}
P(0) & = \left(
\begin{array}{ccccc}
0 & 0 & 0 & 0 & 0 \\
0 & 1 & 0 & 0 & 0 \\
0 & 0 & 0 & 0 & 0 \\
0 & 0 & 0 & 0 & 0 \\
0 & 0 & 0 & 0 & 1 \\
\end{array}\right)~, \\
P(1) & = \left(
\begin{array}{ccccc}
1 & 0 & 0 & 0 & 0 \\
0 & 0 & 0 & 0 & 0 \\
0 & 0 & 1 & 0 & 0 \\
0 & 0 & 0 & 1 & 0 \\
0 & 0 & 0 & 0 & 0 \\
\end{array}\right)~.
\end{align}

The difference between the measurement alphabet size and the period of a
process, which determines the number of states of a periodic process,
should be noted. In all our examples the measurement
alphabet is binary. Thus, in having five internal states but only a two-letter
measurement alphabet, the period-$5$ processes necessarily constitute
systems observed via incomplete measurements.

The set of languages generated by the three processes is summarized in
Table~\ref{tab:p5}. The generated language depends on the initial state only
when the measurement period is a multiple of the process period.

\begin{table}[tbp]
\begin{tabular}{|l|c|c|c|c|}
\hline
\multicolumn{5}{|c|}{Distinct Period-$5$ Processes' Language Diversity} \\
\hline
Machine & $\mperiod$ & $\supp(\Lang)$ & $\Lang, L > 5$ & $\Ldiv$\\
Type & & & & \\
\hline
10000 & $5n+1, 5n+2$ & $\subw((10000)^*)$ & $\Prob(y^L) = 1/5$ & \\
      & $5n+3, 5n+4$ &                    &                 	& \\
      & $5n$         & $0^*$              & $\Prob(y^L) = 1$	& \\
      & $5n$         & $1^*$              & $\Prob(y^L) = 1$	& $1.58$\\
\hline
11000 & $5n+1, 5n+4$ & $\subw((11000)^*)$ & $\Prob(y^L) = 1/5$	& \\
      & $5n+2, 5n+3$ & $\subw((01010)^*)$ & $\Prob(y^L) = 1/5$	& \\
      & $5n$         & $0^*$              & $\Prob(y^L) = 1$	& \\
      & $5n$         & $1^*$              & $\Prob(y^L) = 1$	& $2$ \\
\hline
10101 & $5n+1, 5n+4$ & $\subw((10101)^*)$ & $\Prob(y^L) = 1/5$	& \\
      & $5n+2, 5n+3$ & $\subw((00111)^*)$ & $\Prob(y^L) = 1/5$	& \\
      & $5n$         & $0^*$              & $\Prob(y^L) = 1$	& \\
      & $5n$         & $1^*$              & $\Prob(y^L) = 1$	& $2$ \\
\hline
\end{tabular}
\caption{Process languages produced by the three distinct period-$5$ generators.
  The quantum and classical versions are identical in each case.
  The measurement period is parametrized by $n=0,1,2\dots$.
  For simplicity, the word probability is given for all $y^L\in\Lang$
  with $L \geq 5$. For the nontrivial languages above, when $L > 5$ there
  are only five words at each length, each having equal probability.
  }
\label{tab:p5}
\end{table}

The language diversity for the process $10000$ is $\Ldiv = \log_2 (3)$
and for both the processes $11000$ and $10101$, $\Ldiv = 2$. Note that the
processes $11000$ and $10101$ generate each other at particular measurement
periods, if one exchanges $0$s and $1$s. It is not surprising therefore
that the two models have the same language diversity.

It turns out that the state of the quantum systems under periodic dynamics is
independent of the measurement protocol. At each point in time the system is
 in an eigenstate of the measurement operator. Therefore, the measurement does
not alter the internal state of the quantum system. Thus, a system in state
$\langle\psi_0\vert$ is going to be in a particular state $\langle\psi_2\vert$
after two time steps, independent of whether being measured in between. This is
true for quantum and classical periodic systems.
 The conclusion is that for periodic processes there is no difference between
unmeasured quantum and classical states. This is worth noting, since this is
the circumstance where classical and quantum systems are supposed to differ.
As a consequence the language diversity is the same for the quantum and
classical model of all periodic processes, which coincides with 
Prop.~\ref{the:perstochlang}.

Note, however, that the language diversity is not the same for all processes
with the same period. A property that is reminiscent of the transient
information \cite{crutchfield:03,Feld02a}, which also distinguishes between
structurally different periodic processes.

\subsection{Discussion}

The examples show that the language diversity monitors aspects of a process's
structure and it is different for quantum and classical models of aperiodic
processes. This suggests that it will be a useful aid in discovering structure
in the behavior of quantum dynamical systems.
For the aperiodic examples, the $QDG$ had a larger language diversity than its
classical $DG$. And this suggests a kind of computational power of $QDG$s that
is not obvious from the structural constraints of the machines.
Language diversity could be compensation, though, for other limitations
of $QDG$s, such as not being able to generate all regular languages.
The practical consequences of this for designing quantum devices
remains to be explored.

\begin{table}
\begin{tabular}{|l | cc | cc |}
\hline
				& Quantum process 			& Classical process 		\\
\hline
System			& Iterated beam splitter	& Fair coin					\\
$\Ldiv$			& $\log_2(3)$				& $0$        \\
Measurement 	& Complete					& Complete	\\
\hline
System			& Quantum kicked top 		& Fair coin	\\
$\Ldiv$			& $log_2(5)$			    & $0$        \\
Measurement     & Complete					& Complete	\\
\hline
\hline
System			& 10000						& 10000	\\
$\Ldiv$			& $\log_2(3)$				& $\log_2(3)$	\\
Measurement 	& Incomplete				& Incomplete	\\
\hline
System			& 11000						& 11000	\\
$\Ldiv$			& $2$						& $2$	\\
Measurement		& Incomplete				& Incomplete	\\
\hline
System			& 10101						& 10101	\\
$\Ldiv$			& $2$						& $2$	\\
Measurement     & Incomplete				& Incomplete	\\
\hline
\end{tabular}
\caption{Comparison between $QDG$s and their classical $DG$s. Note that the
term ``(in)complete measurement'' is not used for classical systems. However,
the above formalism does render it meaningful. It is used in the same way as in
the quantum case (one-dimensional  subspaces or non-degenerate eigenvalues).
  }
\label{tab:sum}
\end{table}

A comparison between $QDG$s and their classical $DG$s gives a first hint at the
structure of the lowest levels of a potential hierarchy of quantum computational
model classes. It turned out that for periodic processes a $QDG$ has no
advantage over a $DG$ in terms of the diversity of languages possibly
generated by any $QDG$. However, for the above examples of both incomplete and complete
measurements, the set of generated stochastic languages is larger for a $QDG$
than the corresponding $DG$.

Table~\ref{tab:sum} summarizes the processes discussed above, their
properties and language diversities. All finite-state machines are
deterministic, for which case it was shown that there exists an
equivalent $DG$ that generates the same language \cite{wiesner:06}. This
is true, though only for one particular measurement period. Here we expanded
on those results in comparing a range of measurement periods and the entire
set of generated stochastic languages.

For each example quantum generator and the
corresponding classical generator the language diversity and the type of
measurement (complete/incomplete) are given. For all examples the
language diversity is larger for the $QDG$ than the $DG$. It should be
noted, however, that the fair coin process is also generated by a
one-state $DG$ with transition matrices $T(0) = T(1) = (1/2)$. This it
not true for the $QDG$s. Thus, the higher language diversity of a $QDG$
is obtained at some cost---a larger number of states is needed than with
a $DG$ generating any one particular process language. The situation is
different, again, for the period-5 processes---there is no $DG$ with
fewer states that generates the same process language. 

The above examples were simple in
the sense that their language diversity is a finite, small number. In some
broader sense, this means that they are recurrent---to use
terminology from quantum mechanics. For other processes the situation might
not be quite as straightforward. To find the language diversity one has to
take the limit of large measurement periods.
For implementations this is a trade-off, since larger measurement period
requires a coherent state for a longer time interval. In particular it should
be noted that in the above examples shorter 
intervals between measurements cause more ``interesting'' observed behavior.
That is, the stochastic language $\Lang^2 = \{(01)^*,(10)^*\}$ generated by
the quantum kicked top with $\Prob(y^L)=1/2$, consisting of strings with
alternating $0$s and $1$s is more structured than the language
$\Lang^4 = \{0^*\}$ with $\Prob(y^L) = 1$ consisting of only $0$s.
(Cf. Table~\ref{tab:qkt}.)

\section{Conclusion }

Quantum finite-state machines occupy the lowest level of an as-yet only
partially known hierarchy of quantum computation. Nonetheless, they are
useful models for quantum systems that current experiment can implement,
given the present state of the art. We briefly reviewed quantum finite-state
generators and their classical counterparts---stochastic finite-state
generators. Illustrating our view of computation as an intrinsic property of a
dynamical system, we showed similarities and differences between finite-memory
classical and quantum processes and, more generally, their computational model
classes. In particular, we introduced the language diversity---a new property
that goes beyond the usual comparison of classical and quantum machines.
It captures the fact that, when varying measurement protocols, different
languages are generated by quantum systems. Language diversity appears
when quantum interference operates.

For a set of examples we showed that a deterministic quantum finite-state
generator has a larger language diversity than its classical analog.
Since we associate a language with a particular behavior, we also associate
a set of languages with a set of possible behaviors. As a consequence, the
$QDG$s all exhibited a larger set of behaviors than their classical analogs.
That is, they have a larger capacity to store and process information. 

We close by suggesting that the design of finite quantum computational
elements could benefit from considering the measurement process not only as a
final but also as an intermediate step, which may simplify experimental
design.

Since we considered only finite-memory systems here, their
implementation is already feasible with current technology. Cascading
compositions of finite processes can rapidly lead to quite sophisticated
behaviors, as discussed in Ref.  \cite{wiesner:06}. A discussion of associated
information storage and processing capacity analogous to those used for
classical dynamical systems in Ref. \cite{crutchfield:03} is under way.

\begin{acknowledgments}
Partial support was provided by DARPA Agreement F30602-00-2-0583. KW's
postdoctoral fellowship was provided by the Wenner-Gren Foundations,
Stockholm, Sweden.
\end{acknowledgments}

\bibliography{ref}
\end{document}